\title{The Impact of Digital Financial Services on Firm's Performance:a Literature Review}
\author{
Tariq Abbasi \footnote{To whom correspondence should be addressed; E-mail: tariq.abbasi@nbs.nust.pk} , and Hans Weigand\\ 
                Department of Management\\
        Tilburg School of Economics and Management\\
        Tilburg University, The Netherlands\\
}
\date{\today}

\documentclass[12pt]{article}
\usepackage{float}
\usepackage{hyperref}
\usepackage{adjustbox}
\usepackage[utf8]{inputenc}
\usepackage{tikz}
\usetikzlibrary{shapes,arrows}
\usepackage[top=2.5cm,bottom=2.5cm,left=2cm,right=2cm]{geometry}

\usepackage[round, authoryear,sort,numbers]{natbib}

\begin{document}
\maketitle

\begin{abstract}
Digital Financial Services continue to expand and replace the delivery of traditional banking services to the customers through innovative technologies to meet the growing complex needs and globalization challenges. These diversified digital products help the organizations (service providers) to improve their firm performance and to remain competitive in the market. It also assists in increasing market share to grow their profitability and improve financial position. There is a growing literature on Digital Financial Services and firm performance. At this point of the development, this paper systemically reviews the existing (within last one decade) amount of literature investigating the impact of DFS on firm’s performance, analyzes and identifies the research gaps. We identify 39 works that have appeared in a wide range of peer-reviewed scientific journals. We classify the methodologies and approaches that researchers have used to predict the effect of such services on the financial growth and profitability.  We observe that despite rapid technological advancement in DFS during the last ten years, Digital Financial Services being the factor affecting firm's performance didn't get the reasonable attention in academic literature. One of the reason is that almost all the authors limit their research to banking sector while ignoring others particularly mobile network operators (providing branchless banking) and new non-banking entrants. We also notice that newer researchers often ignore past research and investigate the same issues. This study also makes several recommendations and suggest directions for future research in this still emerging field. 
\end{abstract}


\section{Introduction}\label{intro}
The revolution of information technology (IT) has changed every aspect of human being’s life including banking. IT works as a catalyst for growth in the banking sector; particularly it supports banking services, productivity growth, and risk management. IT is a business driver and thus can be used to reinforce competitive edge \citep{Porter:1985}. From the last one decade, financial institutions invested heavily in building their IT infrastructure which enabled them a successful transformation towards digital banking. Firms investment in IT is under the presumption that such investments would enhance operating efficiency and thus improve financial performance.\\

According to \citep{Lodge:2016} banks in North America, Europe, Asia-Pacific and Latin America will together spend \$241bn on IT in 2016, with an overall increase of almost 4\% compared with 2015. We can divide these IT investments theoretically in two parts i-e internal and external purposes. The internal purpose we mean the investments for risk management, regulatory, productivity and business models. External purpose IT investments are solely for the digital financial services. The later part is more important from the perspective of a customer who is the user of such services. Thus, the issue of service marketing in general and particularly digital banking services witnessed a substantial growth during the last decade in almost all the countries around the world.\\

Digital financial services (DFS) expand the delivery of traditional banking services to the customers through innovative technologies like internet banking, mobile-phone-enabled solutions, electronic money models and digital payment platforms. Although modern digital banking started with the automated teller machines (ATM) and phone banking, however, the internet and mobile banking offer fast and effective delivery channels not only for traditional banking products but also paved the way for new products as well. The outreach of 3G and 4G internet technology along with the expanded uses of smartphones and tablets has increased the demand for digital services. This market demand encourages financial institutions, software houses and other service providers to offer advanced digital banking services together with the advent of new diversified products and applications to retain the existing clients and access the unbanked population. \\

According to the annual Measuring the Information Society (ITU’s) Report at the end of 2016, there are almost as many mobile-cellular subscriptions as people on earth and 95\% of the global population lives in an area that is covered by a mobile cellular signal. Also by the end of 2016, 47\% of the global population is online—which is certainly progress, but also means that over half of the world’s population, or roughly 3.9 billion people, is not. The ITU’s predict online access in 2020 will reach 60\% of the world’s population. As per leading Fintech analysts, Juniper Research, over 2bn mobile users will have used their devices for banking purposes by the end of 2021, compared to 1.2bn 2016 globally. These trends suggest potential growth opportunities remain, leading to predictions of massive increases in the number of DFS users. \\

Access to financial services can serve as a bridge out of poverty. The number of people worldwide having an account grew by 700 million between 2011 and 2014. 62 percent of the world’s adult population has an account; up from 51 percent in 2011. Three years ago, 2.5 billion adults were unbanked. Today, 2 billion adults remain without an account. This represents a 20 percent decrease \citep{Asli:2015}. This shows the growth of DFS through different modes (mostly mobile and internet banking) and the potential to achieve the goal of universal financial access.\\
 
From the service provider perspective “'there ain't no such thing as a free lunch”. Their motive is to improve their firm performance through enhance operational efficiency, increase market share, innovative products, extend their client reach, improve customer retention, new employment opportunities and software applications etc. to progress and to remain competitive in the market. \\

The banking industry (now also newcomers like IT companies) is one of the pioneers of IT-based financial services applications and today heavily depends on DFS in reshaping its service delivery systems. The first ATM was installed in the late 1960s, several DFS have already been utilized for years by the banking industry since then. \\

Many studies analyze the effect of DFS on the firm performance, using both quantitative and qualitative methodologies. Despite considerable research on DFS and firm performance that has appeared in academic journals, a review of the literature on this topic remains missing. Such reviews play an important role in the development of research field through identification of areas where research is lacking. Reviewing existing literature not only provides an opportunity to better understand the current state of the research, but it also distinguishes patterns in the progress of the research area itself. The aim of this study is to explore, understand, analyze and summarize findings of DFS on firm performance in the last one decade (2006-2016). It effectively accumulates the existing body of research into the more coherent body of knowledge, and to interpret previous findings. It also identifies and discusses the models and frameworks applied so far in this area. Summarizes the major findings and discover important gaps which need more attention and demands further research. \\


\section{Digital Financial Services}\label{DFS}
DFS are the broad range of financial services accessed and delivered through digital channels, including payments, credit, savings, remittances, insurance and financial information. The term “digital channels” refers to the internet, mobile phones (both smartphones and digital feature phones), ATMs, POS terminals, NFC-enabled devices, chips, electronically enabled cards, biometric devices, tablets, phablets and any other digital system \citep{AFI:2016}. \\

DFS have significant potential to expand the delivery of basic financial services via affordable, convenient and secure environment to the public at large (particularly the poor) through innovative technologies like mobile-phone-enabled solutions, electronic money models, and digital payment platforms. Financial Institutions (Banks, Microfinance institutions) and non-Financial firms (mobile network operators) and third party providers (agent network managers, payment aggregators, and others) are leveraging digital channels to offer basic financial services at greater convenience, scale and lower cost than traditional banking allow. The digital revolution adds new layers to the material cultures of financial inclusion, offering the state new ways of expanding the inclusion of the ‘legible’, and global finance new forms of ‘profiling’ poor households into generators of financial assets \citep{Gabor:2016}.\\

Mobile telephony (DFS) exerts a significant and non-negligible impact on economic growth \citep{Saibal:2016}.The extensive use of digital finance could boost annual GDP of all emerging economies by \$3.7 trillion by 2025 (6\% increase). Approximately 65\% of this increase would come from raised productivity of financial and non-financial businesses and governments as a result of digital payments. This additional GDP could lead to the creation of up to 95 million jobs across all sectors \citep{James:2016}.\\

As DFS include many digital channels (improving day to day).We noticed that in our identified articles (2006 to 2016) researchers mainly focus on one particular channel at a point in time to execute their research (see Table\ref{table1a}). Most of the studies used internet banking (34\%) as DFS channel followed by automated/internet/self-service technology (32\%). 
\begin{table}[H]
\centering
\caption{DFS used in literature review}
\begin{adjustbox}{width=1\textwidth}
\label{table1a}
\begin{tabular}{|l|l|l|}

\hline\hline
No & Nomenclature used for digital channel               & Author                                                                                          \\
\hline\hline
           &                                        &                                                                                                 \\
1 & e-banking                                          & \citep{Mohammad:2011}, \citep{SAEID:2014}, \citep{Abaenewe:2013}  \\
  & 												 &  \citep{Ugwueze:2016},\citep{Akhisar:2015},\citep{Siddik:2016}\\  
	&											    &                                                                                           \\
2 & internet banking                                   & \citep{Pooja:2006},\citep{Ceylan:2008}, \citep{Kennedy:2013}                                     \\
   & 											   & \citep{Pooja:2009},   \citep{Van:2015},\citep{Pooja:2010}\\
	&										       & \citep{Karimzadeh:2013},\citep{Lin:2011},\citep{Tunay:2015}	\\
	&										       & \citep{Stoica:2015},  \citep{Dandapani:2008},\citep{Delgado:2007}\\
	&										       & \citep{Callaway:2011}  \\         
											   
	&										        &                                                                                           \\
3 & mobile banking                                     & \citep{Kennedy:2013}                                                                          \\
	&											 &                                                                                                 \\
4 & online banking                                     & \citep{Wu:2010},\citep{Acharya:2008},\citep{Ho:2009}                                      \\

		&											 &                                                                                                 \\
 5 & cash less banking                                  & \citep{KASHIF:2016}                                                                           \\
		
&                                                      &                                                                            \\
6 & automation/internet/                                & \citep{BILAL:2015},\citep{Abubakar:2013},\citep{Georgia:2013}\\
  & self-service technology							& \citep{Ciciretti:2009}, \citep{Uchida:2011},\citep{AlHawari:2006}\\
	&												&   \citep{Hernando:2007},\citep{Meepadung:2009},\citep{DeYoung:2007}\\ 
	&												&   \citep{Hung:2012},\citep{Lavinia:2014},\citep{Del-Giudice:2016}\\
		&                                                &						                                                                                          \\
 7 & ATM                                                & \citep{Georgia:2015},\citep{Kondo:2010}  \\                                                  

\hline
\end{tabular}
\end{adjustbox}
\end{table}

\section{Research Methodology}\label{RM}
                                                                
To determine the current state of research pertaining to DFS and firm performance we conducted an extensive literature review. Since DFS is an interdisciplinary topic hence the relevant research articles are published in a wide variety of academic journals and few records of conference proceedings. To identify the relevant articles, we have done a systemic scan of various databases (academic journals \& Conferences) through multiple relevant key terms (see Table\ref{table1}).
\begin{table}[H]
\centering
\caption{Databases and key terms}
\begin{adjustbox}{width=1\textwidth}
\label{table1}
\begin{tabular}{|l|l|l|}
\hline\hline 

              Sources             & Objective Key terms               & Subjective key terms      \\
                           \hline\hline
                           
                                    &                                   &                           \\
·      Science Direct               & ·      Digital Financial Services & ·      Firm performance   \\
·      Emerald                      & ·      Mobile banking             & ·      Firm profitability \\
·      ProQuest Direct              & ·      Internet banking           & ·      Firm efficiency    \\
·      Taylor \& Francis            & ·      Mobile payments            & ·      Financial impact   \\
·      Wiley                        & ·      Branchless banking         & ·      Cost effective     \\
·      Inderscience                 & ·      Online banking             & ·      Bank performance   \\
·      ACM Digital Library          & ·      Electronic banking         &                           \\
·      AIS Electronic Library       & ·      Digital banking            &                           \\
·      IEEE- Xplore Digital Library & ·      Mobile money               &                           \\
·      Scopus                       &                                   &                           \\
·      Springer                     &                                   &                           \\
·      Google Scholar               &                                   &                          \\
\hline
\end{tabular}
\end{adjustbox}
\end{table}
To find a relevant article in a database each key objective term is searched with the combination of subjective terms. Through identified articles, we explored the citations and references to find any previous or leading work that is in our scope. In addition, Google scholar identified certain articles which were not accessible in the other databases mentioned in table1.  Initially 52 articles were found, however, a two-step pre-selection is performed before including the article in our literature review. In the first step, we discard few articles for which the publisher or journal is identified as unauthenticated. This step is necessary as we used Google scholar as well other than scientific databases. In the second step, abstracts are read to eliminate the articles that do not deal with the subject. \\
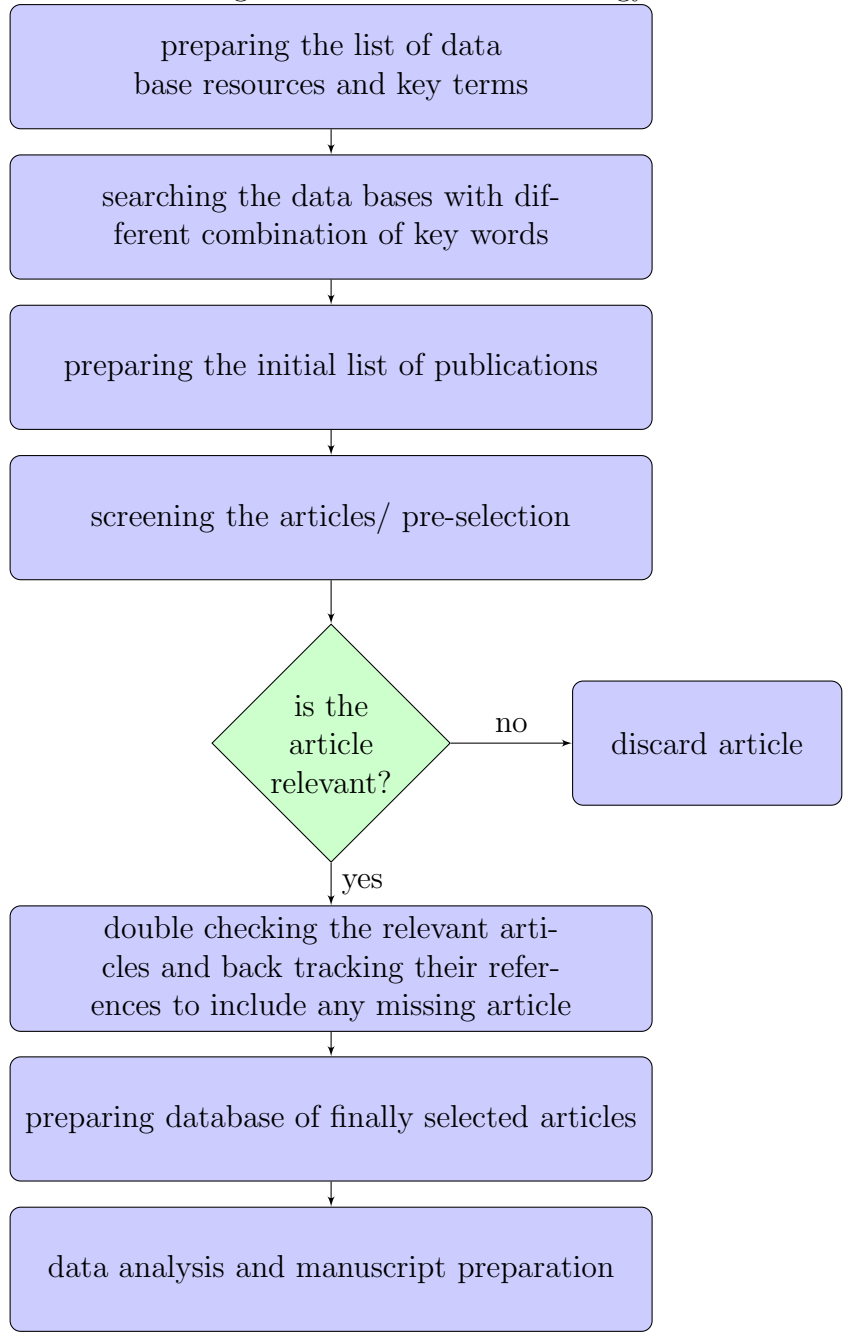
\begin{figure}[H]
\centering
\caption{Research Methodology}

\tikzstyle{decision} = [diamond, draw, fill=green!20, 
    text width=4.5em, text badly centered, node distance=3cm, inner sep=0pt]
\tikzstyle{block} = [rectangle, draw, fill=blue!20, 
    text width=20em, text centered, rounded corners, minimum height=4em]
    \tikzstyle{block1} = [rectangle, draw, fill=blue!20, 
    text width=08em, text centered, rounded corners, minimum height=4em]
\tikzstyle{line} = [draw, -latex']
    
\begin{tikzpicture}[node distance = 2cm, auto]
    \node [block] (init) {preparing the list of data base resources and key terms};
    \node [block, below of=init] (identify) {searching the data bases with different combination of key words};
    \node [block, below of=identify] (evaluate) {preparing the initial list of publications};
   \node [block, below of=evaluate] (evaluates) {screening the articles/ pre-selection};
    \node [decision, below of=evaluates] (decide) {is the article relevant?};
     \node [block1, right of=decide, node distance=5cm] (update) {discard article};
    \node [block, below of=decide, node distance=3cm] (stop) {double checking the relevant articles and back tracking their references to include any missing article};
    \node [block, below of=stop, node distance=2cm] (stop1) {preparing database of finally selected articles};
    \node [block, below of=stop1, node distance=2cm] (stop2) {data analysis and manuscript preparation};
   
    \path [line] (init) -- (identify);
    \path [line] (identify) -- (evaluate);
    \path [line] (evaluate) -- (evaluates);
    \path [line] (evaluates) -- (decide);
    \path [line] (decide) -- node  {no} (update);
    \path [line] (decide) -- node {yes}(stop);
    \path [line] (stop) -- (stop1);
    \path [line] (stop1) -- (stop2);
\end{tikzpicture}
\end{figure}
The final selection resulted in 39 relevant research articles, wherein 36 were published in scientific journals and 2 in conferences and 1 in SSRN.  Hence the selected articles represented a wide range of peer-reviewed scientific journals. The period reviewed is from January 2006-December 2016 (Inclusive). The period of last one decade is selected mainly due to two reasons. Firstly, the companies (particularly banking) started IT investments in the 90s but heavily invested in DFS (online banking) products from the year 2000 onwards. Secondly, due to the sharp growth in mobile and internet accessibility from the year 2005 onwards the usage of DFS improve and continue to increase.The companies transform from traditional to the digital model in this era to fill the customer needs, remain in the competition, improve efficiency and hence to reap the benefits of digitization. So we decided to review the research of last one decade. The selected articles used different research approaches and methodologies. The studies were conducted in different regions including both developed and developing countries.    
\begin{table}[H]
\centering
\caption{Articles on DFS and Firm financial performance (Jan 2006-Dec 2016)}
\begin{adjustbox}{width=1\textwidth}

\label{table2}
\begin{tabular}{|l|l|llllllllllll|l|}
\hline\hline 
No & Journal/Conference Name                                                       & 2006 & 2007 & 2008 & 2009 & 2010 & 2011 & 2012 & 2013 & 2014 & 2015 & 2016 & Total & \%    \\
\hline\hline
   &                                                                               &      &      &      &      &      &      &      &      &      &      &      &       &       \\
1  & Banks and Bank Systems                                                        & 1    & -    & -    & -    & -    & -    & -    & -    & -    & -    & -    & 1     & 3\%   \\
2  & Marketing Intelligence \& Planning                                            & 1    & -    & -    & -    & -    & -    & -    & -    & -    & -    & -    & 1     & 3\%   \\
3  & South Asian Journal of Management                                             & 1    & -    & -    & -    & -    & -    & -    & -    & -    & -    & -    & 1     & 3\%   \\
4  & Journal of Banking \& Finance                                                 & -    & 2    & -    & -    & -    & -    & -    & -    & -    & -    & -    & 2     & 5\%   \\
5  & European Financial Management                                                 & -    & 1    & -    & -    & -    & -    & -    & -    & -    & -    & -    & 1     & 3\%   \\
6  & Oxford Business \&Economics Conference Program                                & -    & -    & 1    & -    & -    & -    & -    & -    & -    & -    & -    & 1     & 3\%   \\
7  & International Journal of Bank Marketing                                       & -    & -    & 1    & -    & -    & -    & -    & -    & -    & -    & -    & 1     & 3\%   \\
8  & Managerial Finance                                                            & -    & -    & 1    & -    & -    & -    & -    & -    & -    & -    & -    & 1     & 3\%   \\
9  & Eurasian Journal of Business and Economics                                    & -    & -    & -    & 1    & -    & -    & -    & -    & -    & -    & -    & 1     & 3\%   \\
10 & Computers \& Operations Research                                              & -    & -    & -    & 1    & -    & -    & -    & -    & -    & -    & -    & 1     & 3\%   \\
11 & Journal of High Technology Management Research                                & -    & -    & -    & 1    & -    & -    & -    & -    & -    & -    & -    & 1     & 3\%   \\
12 & Journal of Financial Services Research                                        & -    & -    & -    & 1    & -    & 1    & -    & -    & -    & -    & -    & 2     & 5\%   \\
13 & Int. J. Electronic Finance                                                    & -    & -    & -    & -    & 1    & -    & -    & -    & -    & -    & -    & 1     & 3\%   \\
14 & Kybernetes                                                                    & -    & -    & -    & -    & 1    & -    & -    & -    & -    & -    & -    & 1     & 3\%   \\
15 & Applied Economics Letters                                                     & -    & -    & -    & -    & 1    & -    & -    & -    & -    & -    & -    & 1     & 3\%   \\
16 & Journal of Engineering and Technology Management                              & -    & -    & -    & -    & -    & -    & 1    & -    & -    & -    & -    & 1     & 3\%   \\
17 & European Scientific Journal                                                   & -    & -    & -    & -    & -    & -    & -    & 2    & -    & -    & -    & 2     & 5\%   \\
18 & West African Journal of Industrial and Academic Research                      & -    & -    & -    & -    & -    & -    & -    & 1    & -    & -    & -    & 1     & 3\%   \\
19 & Acta Universitatis Danubius. Œconomica                                        & -    & -    & -    & -    & -    & -    & -    & 1    & -    & -    & -    & 1     & 3\%   \\
20 & Int. J. Financial Services Management                                         & -    & -    & -    & -    & -    & -    & -    & 1    & -    & -    & -    & 1     & 3\%   \\
21 & Procedia - Social and Behavioral Sciences                                     & -    & -    & -    & -    & -    & -    & -    & -    & -    & 2    & -    & 2     & 5\%   \\
22 & Pakistan Economic and Social Review                                           & -    & -    & -    & -    & -    & -    & -    & -    & -    & 1    & -    & 1     & 3\%   \\
23 & Procedia Economics and Finance                                                & -    & -    & -    & -    & -    & -    & -    & -    & -    & 1    & -    & 1     & 3\%   \\
24 & Global Business and Economics Review                                          & -    & -    & -    & -    & -    & -    & -    & -    & -    & 1    & -    & 1     & 3\%   \\
25 & Journal of Internet Banking \& Commerce                                       & -    & -    & -    & -    & -    & 1    & -    & -    & -    & 1    & -    & 2     & 5\%   \\
26 & Paradigms: A Research Journal of Commerce, Economics, and Social Sciences     & -    & -    & -    & -    & -    & -    & -    & -    & -    & -    & 1    & 1     & 3\%   \\
27 & Journal of Business Economics and Management                                  & -    & -    & -    & -    & -    & -    & -    & -    & -    & -    & 1    & 1     & 3\%   \\
28 & SSRN                                                                          & -    & -    & -    & -    & -    & 1    & -    & -    & -    & -    & -    & 1     & 3\%   \\
29 & American Journal of Business                                                  & -    & -    & -    & -    & -    & 1    & -    & -    & -    & -    & -    & 1     & 3\%   \\
30 & Annual Review of Economics                                                    & -    & -    & -    & -    & -    & 1    & -    & -    & -    & -    & -    & 1     & 3\%   \\
31 & 12th International Academic Conference, Prague                                & -    & -    & -    & -    & -    & -    & -    & -    & 1    & -    & -    & 1     & 3\%   \\
32 & Indian Journal of Scientific Research                                         & -    & -    & -    & -    & -    & -    & -    & -    & 1    & -    & -    & 1     & 3\%   \\
33 & International Journal of e-Education, e-Business, e-Management and e-Learning & -    & -    & -    & -    & -    & -    & -    & -    &      & -    & 1    & 1     & 3\%   \\
34 & Business Process Management Journal                                           & -    & -    & -    & -    & -    & -    & -    & -    &      & -    & 1    & 1     & 3\%   \\
   &                                                                               &      &      &      &      &      &      &      &      &      &      &      &       &       \\
   & Total                                                                         & 3    & 3    & 3    & 4    & 3    & 5    & 1    & 5    & 2    & 6    & 4    & 39    & 100\%\\
   \hline

\end{tabular}

\end{adjustbox}
 
\end{table}
\section{Results}\label{concl}
Out of 39 studies included in this literature review, more than half (59\%) of articles were published from 2011-2016 and the rest 41\% were from 2006 to 2010. In 2012 there was only one study. In total 33 journals published articles related to DFS and firm financial performance. Of which 26 published each one article and five published two articles each from 2006 to 2016 (see Table\ref {table2}). \\

Most studies are empirical in nature which is understandable from the title of article wherein topic financial performance is reviewed in the literature. 97\% of the selected articles in this review used the quantitative technique(s) while only one study \citep{Kennedy:2013} was found qualitative in nature. Of 39 selected articles 60\% were published by top-ranked publishers like Elsevier, Emerald, Inderscience, Taylor \& Francis, Wiley, and Springer.  Further, the Elsevier published the most articles (8, 21\%) followed by Emerald (6, 6\%). \\

The studies applied different financial, econometric, mathematical and operation research techniques to investigate the impact of different DFS on the firm’s performance. Regression analysis is the most popular estimating technique which is used in more than 60\% studies. Interestingly all the studies were performed in the banking sector including community and primarily internet banks. The research articles used the number of banks or their branches as sample size. The average sample size was 534, with the largest number of banks in the USA followed by European countries.The maximum sample size (6566) is used by \citep{Dandapani:2008}, while \citep{Mart:2006} and \citep{SAEID:2014} investigated only one bank each.   Geographically the most investigated regions were Europe followed by USA and south Asia (India, Pakistan and Bangladesh). Out of 39 studies 21 (54\%) were conducted in developed countries and the remaining 17 (44\%) in developing country, while one study incorporates both developing and developed countries (see Table\ref{table3}).  \\

\begin{table}[H]
\centering
\caption{Articles of literature review }
\begin{adjustbox}{width=1\textwidth}
\label{table3}
\begin{tabular}{|l|l|l|l|l|}
\hline\hline 
No & Author,Year                & methodology/approach used                                                      & sample size & countries of study \\
  \hline
\hline
   &                            &                                                                                &             &                    \\
   
1  & \citep{Dandapani:2008}   & regression analysis for bank specific financial variables                      & 6566        & USA                \\
2  & \citep{DeYoung:2007}     & percentage change in income and balance sheet selected items                    & 5599        & USA                \\
3  & \citep{Del-Giudice:2016} & classification analysis method (classification and regression tree).           & 3692        & Europe             \\
4  & \citep{Lin:2011}         & propensity-score matching and difference-in-differences methods                & 2487        & USA                \\
5  & \citep{Acharya:2008}     & structural equation modeling and multiple regression analysis                  & 640         & USA                \\
6  & \citep{Callaway:2011}    & regression analysis                                                            & 369         & Turkey             \\
7  & \citep{Hung:2012}        & regression and sensitivity analysis                                            & 181         & Taiwan             \\
8  & \citep{Meepadung:2009}   & data envelopment analysis                                                      & 165         & Thailand            \\
9  & \citep{Kondo:2010}       & regression analysis                                                            & 128         & Japan              \\
10 & \citep{Ciciretti:2009}   & financial ratio  and regression and robust analysis                            & 105         & Italy              \\
11 & \citep{Pooja:2006}       & regression analysis                                                            & 88          & India              \\
12 & \citep{Pooja:2009}       & financial ratio analysis and multivariate analysis                             & 85          & India              \\
13 & \citep{Pooja:2010}       & Financial ratio analysis and multivariate analysis                             & 82          & India              \\
14 & \citep{Akhisar:2015}     & regression analysis                                                            & 82          & 23 countries       \\
15 & \citep{Hernando:2007}    & Financial ratio and multivariate analysis                                      & 72          & Spain              \\
16 & \citep{Delgado:2007}     & Financial ratio  Analysis and The regression analysis of scale and experience  & 60          & Europe             \\
17 & \citep{Ugwueze:2016}     & co-integration and causality approach                                           & 60          & Nigeria            \\
18 & \citep{Uchida:2011}      & financial ratio analysis                                                       & 40          & Bangladesh         \\
19 & \citep{Ho:2009}          & data envelopment and principal component analysis                              & 32          & Taiwan             \\
20 & \citep{Kennedy:2013}     & qualitative analysis                                                           & 30          & Kenya              \\
21 & \citep{BILAL:2015}       & seemingly unrelated regression estimation and cobb-douglas production function & 30          & Pakistan           \\
22 &  \citep{Tunay:2015}    & demitrescu-Hurlin panel causality tests                                        & 30          & Turkey             \\
23 & \citep{KASHIF:2016}      & regression analysis                                                            & 30          & Pakistan           \\
24 & \citep{Stoica:2015}      & data envelopment and principal component analysis                              & 24          & Romania            \\
25 & \citep{Van:2015}         & regression analysis                                                            & 20          & Vietnam            \\
26 & \citep{Ceylan:2011}      & regression analysis                                                            & 18          & Turkey             \\
27 & \citep{Mohammad:2011}    & regression analysis                                                            & 15          & Jordan             \\
28 & \citep{Ceylan:2008}      & regression analysis for bank specific financial variables                      & 14          & Turkey             \\
29 & \citep{Siddik:2016}      & regression analysis                                                            & 13          & Bangladesh         \\
30 & \citep{Abubakar:2013}    & regression analysis                                                            & 11          & NIgeria            \\
31 & \citep{Georgia:2013}     & regression analysis                                                            & 11          & Greece             \\
32 & \citep{Lavinia:2014}     & regression analysis                                                            & 11          & Romania            \\
33 & \citep{Georgia:2015}     & data envelopment and regression analysis                                       & 11          & Greece             \\
34 & \citep{Wu:2010}          & data envelopment and principal component analysis                              & 10          & USA \& UK          \\
35 & \citep{Karimzadeh:2013}  & regression analysis                                                            & 8           & India              \\
36 & \citep{AlHawari:2006}    & effect of service quality on bank profitability: structural equation modeling & 4           & Australia          \\
37 & \citep{Abaenewe:2013}    & financial ratio analysis                                                       & 4           & Nigeria            \\
38 & \citep{Mart:2006}        & activity-based-costing (ABC)                                                   & 1           & Estonia            \\
39 & \citep{SAEID:2014}       & regression analysis                                                            & 1           & Iran              \\\hline
\end{tabular}
\end{adjustbox}
\end{table}
We also list the 15 most-cited articles from our literature review (see Table\ref{table4}). We made the cut at 20 citations from Google Scholar as the citation for some articles was very less. These 15 studies were mainly conducted in developed countries with exception of two articles in India and Kenya. The articles in journals with higher impact factors or published by the high ranked publishers are the most-cited (citation above 100) publications. Out of 39 articles, only four publications could cross the mark of 100 citations so far. These 15 most cited articles filled the research gap, provide definitions of financial digital delivery channels, a comprehensive methodology with suitable sample size, references and paved the way for future research. \\
\begin{table}[H]
\centering
\caption{15 most cited articles}
\begin{adjustbox}{width=1\textwidth}
\label{table4}
\begin{tabular}{|l|l|l|l|l|}
\hline\hline
No & Title                                                                                                                             & Journal/Conference                             & year & citation \\
\hline\hline

1  & The effect of automated service quality on Australian banks' financial performance and the mediating role of customer satisfaction & Marketing Intelligence \& Planning             & 2006 & 241      \\
2  & How the Internet affects output and performance at community banks                                                                & Journal of Banking \& Finance                  & 2007 & 199      \\
3  & Is the Internet delivery channel changing banks’ performance? The case of Spanish banks                                            & Journal of Banking \& Finance                  & 2007 & 176      \\
4  & Online banking performance evaluation using data envelopment analysis and principal component analysis                             & Computers \& Operations Research               & 2009 & 105      \\
5  & The Impact of Internet Banking on Bank Performance and Risk: The Indian Experience                                                & Eurasian Journal of Business and Economics     & 2009 & 83       \\
6  & Do European primarily internet banks show scale and experience efficiencies?                                                       & European Financial Management                  & 2007 & 73       \\
7  & Do Internet Activities Add Value? Evidence from the Traditional Banks                                                              & Journal of Financial Services Research         & 2009 & 48       \\
8  & The impact of e-banking on the performance of Jordanian banks.                                                                    & Journal of Internet Banking \& Commerce        & 2011 & 47       \\
9  & The impact of mobile and internet banking on performance of financial institutions in Kenya                                       & European Scientific Journal                    & 2013 & 47       \\
10 & Online banking applications and community bank performance                                                                        & International Journal of Bank Marketing        & 2008 & 44       \\
11 & Internet banking services and credit union performance                                                                            & Managerial Finance                             & 2008 & 35       \\
12 & The impact of Internet-Banking on Bank Profitability- The Case of Turkey                                                          & Oxford Business \&Economics Conference Program & 2008 & 29       \\
13 & The impact of Internet banking on bank’s performance: The Indian experience                                                       & South Asian Journal of Management,             & 2006 & 27       \\
14 & IT-based banking services: Evaluating operating and profit efficiency at bank branches                                            & Journal of High Technology Management Research & 2009 & 26       \\
15 & Performance evaluation and risk analysis of online banking service                                                                & Kybernetes                                     & 2010 & 21 \\
\hline     
\end{tabular}
\end{adjustbox}
\end{table}
In Table\ref{table5}, we gathered research techniques used to investigate the impact of different DFS on the firm’s performance. In our literature review (39 articles) we could find eight different mathematical, financial, econometric, and operation research techniques. Most of the studies (70\%) combined financial ratios and regression analysis followed by (13\%) data envelopment and principal component analysis. In addition, As Table\ref{table5} reveals, few authors (e.g \citep{Mart:2006},\citep{Del-Giudice:2016},\citep{BILAL:2015},\citep{Lin:2011},) used one specific technique which is not used by any other author in the entire sample of review.  \\ 

\begin{table}[H]
\centering
\caption{List of techniques used}
\begin{adjustbox}{width=1\textwidth}
\label{table5}
\begin{tabular}{|l|l|l|}
\hline\hline 
No & methodology /technique used                                                            & \# of articles \\
\hline\hline 
   &                                                                                       &                \\
1  & Financial ratio, percentage change (income and balance sheet) and Regression Analysis & 27             \\
2  & Data envelopment and principal component analysis                                     & 5              \\
3  & Structural equation modeling                                                         & 2              \\
4  & Co-integration and causality approach                                                  & 2              \\
5  & Activity-based-costing (ABC)                                                          & 1              \\
6  & Propensity-score matching and difference-in-differences methods                       & 1              \\
7  & Cobb-Douglas production function                                                      & 1              \\
8  & Classification analysis method                                                        & 1             \\
\hline
\end{tabular}
\end{adjustbox}
\end{table}
We identified different dependent and independent variables used in the prior research to unfold the impact of DFS on the firm’s performance. As many as  28 dependent and 16 independent variables have been identified (see Table\ref{table6}). Among the identified dependent variables, ROE and ROA are the most commonly used; approximately two-third of all studies cite them. The other more frequently used dependent variables include income (net, non-interest, and interest), deposits, operating expense, staff costs, efficiency ratio and average internet users. More than half (15) of dependent variables, as listed in Table\ref{table6}, have been tested only once or twice. Out of 16 independent variables used to inspect the impact of DFS on firm's performance, the most frequently used are; the number of ATMs, internet adoption, and internet banking. Notably, one-third of all reviewed articles cite them. The other more frequently used independent variables are the number of branches, POS, the number of employees, cards (debit or credit), website, electronic and mobile banking. Approximately One-third of all independent variables, as listed in Table\ref{table6}, have been tested only once or twice.\\

In order to control the effect of other variables on the financial performance at the same time while examining the DFS relationship, many studies used different control variables. These include financial (bank size, loans, deposits etc) and economic (GDP, inflation, job growth etc), as listed in Table\ref{table6}. \\

\begin{table}[]
\centering
\caption{Dependent, Independent and control variables}
\begin{adjustbox}{width=1\textwidth}
\label{table6}
\begin{tabular}{|l|llllllll|}
\hline\hline
No & Dependent Variable                              & Frequency &  & Independent variable           & Frequency &  & Control Variable & Frequency \\
\hline\hline
   &                                                 &           &  &                                &           &  &                              &           \\
1  & return on equity (ROE)                          & 25        &  & number of ATMs                 & 12        &  & bank size                    & 15        \\
2  & return on assets (ROA)                          & 24        &  & internet adoption              & 11        &  & loans                        & 11        \\
3  & revenue or net income                           & 9         &  & online or internet banking     & 10        &  & deposits                     & 11        \\
4  & non interest income                              & 8         &  & number of branches             & 7         &  & equity                       & 6         \\
5  & interest income                                 & 5         &  & point of sale (POS)            & 5         &  & Gross domestic product (GDP) & 6         \\
6  & domestic deposits                               & 4         &  & number of employees            & 5         &  & inflation                    & 6         \\
7  & growth in assets                                & 4         &  & credit cards                   & 4         &  & operating cost               & 4         \\
8  & operating expense                               & 3         &  & debit cards                    & 4         &  & return on assets (ROA)       & 2         \\
9  & non interest expense to earning assets          & 3         &  & website                        & 4         &  & job growth                   & 2         \\
10 & staff costs                                     & 3         &  & electronic banking             & 3         &  & market share                 & 2         \\
11 & daily reach rate/average internet users         & 3         &  & mobile banking                 & 3         &  & non interest income          & 2         \\
12 & net lending to total assets                     & 3         &  & IT equipment                   & 2         &  & return on equity (ROE)       & 1         \\
13 & efficiency ratio                                & 3         &  & Bank location                  & 2         &  &                              &           \\
14 & IT costs                                        & 2         &  & telephone /Call Center Banking & 2         &  &                              &           \\
15 & marketing expenses                              & 2         &  & IT expense                     & 1         &  &                              &           \\
16 & interest spread                                 & 2         &  & Customer Service Quality       & 1         &  &                              &           \\
17 & total commission and fees to assets              & 2         &  &                                &           &  &                              &           \\
18 & earning per share (EPS)                         & 2         &  &                                &           &  &                              &           \\
19 & global deposits                                 & 1         &  &                                &           &  &                              &           \\
20 & non performing loans to asset ratios            & 1         &  &                                &           &  &                              &           \\
21 & non performing assets to net advances           & 1         &  &                                &           &  &                              &           \\
22 & equity to liabilities                           & 1         &  &                                &           &  &                              &           \\
23 & net interest financial margin to earning assets & 1         &  &                                &           &  &                              &           \\
24 & price to earning ratio                          & 1         &  &                                &           &  &                              &           \\
25 & profit per branch                               & 1         &  &                                &           &  &                              &           \\
26 & equity to total assets                           & 1         &  &                                &           &  &                              &           \\
27 & asset utilization                               & 1         &  &                                &           &  &                              &           \\
28 & willingness to improve level of business        & 1         &  &                                &           &  &                              &          \\
\hline
\end{tabular}
\end{adjustbox}
\end{table}
\section{Research gaps}\label{gaps}
A review of prior, relevant literature is only effective if it creates a firm foundation for advancing knowledge, closes the outdated research areas, identifying the possible research gaps while uncovering areas where future research is needed \citep{Webster:2002}. Given the findings in this literature review, we also outlined certain areas of research where future research should be aiming at.\\
 
 Most of the studies included in this literature review are of quantitative in nature. However, it has been observed that data pertaining to the digital channels used as DFS is either very limited or not empirical in nature. Although financial data indicating the performance of the company is available easily, however, secondary data for digital payment channels were non-existent in the early stage (the 2000s) of emerging DFS. But nowadays availability of resources increased significantly. Hence to improve the quality and applicability of their studies, researchers are expected to use better quality data to validate their research. \\
 
 IT is no more a source of enjoyment for developed countries only. The developing and under developing countries in the recent years (last one decade) invested a lot to develop the IT infrastructure. Both the traditional and digital banking services improved and increased. Since last one-century crossing borders have proved to be difficult both in developed and developing countries. Hence the DFSs varied from country to country with exception of few products e.g globally accepted credit cards. Few of the main reasons for variation are regulations, social and cultural environment and market structure. Therefore, it is difficult to generalize the results of many DFS research. In our literature review we could only find four (\citep{DeYoung:2007},\citep{Akhisar:2015},\citep{Delgado:2007},\citep{Wu:2010}) multi-country studies. Future research covering different countries with comparative market structure could improve the generalizations. A comparative analysis of DFSs performance and its effect on firm with respect to customer based in rural and urban areas could be interesting as well.\\

  The impact of DFSs is no more limited to the banking industry. It is widely being used by other industries as the payment solution for various purposes e.g online shopping, grocery stores, supermarkets, insurance, mobile service providers and trading etc. The 39 research articles included in this review studied the impact of DFS on banking industry (as a service provider) only. We encourage to include more industries e.g IT, Telecom, Energy, Healthcare, Automotive, and Supply Chain companies (as service providers or co-providers) to see how are they getting benefit from such customer oriented services. This more representative sample would provide valuable insights into performance and extend understanding. \\

So far, the research is focused on digital currencies being used for digital channels in DFS. However, cryptocurrencies (e.g bitcoin, ether, monero, Zcash and dash etc) can also use the same digital channels in DFS. Future research on how cryptocurrencies will affect the performance of offering and accepting firm could be interesting. The comparison of crypto versus digital currency for the same accepting firm could also provide interesting results to understand the performance. \\
 
 The branchless banking e.g M–Pesa,  Easypaisa, G-Cash, Ready-cash etc (in developing) and Paypal, skrill, Paysafecard and  Neteller etc (in developed countries), are serving the masses and now a major competitor for traditional banking model. In developing countries, such services are being provided by mobile service providers while in developed by nonbank companies. The future research to investigate the role of such services in the profitability and performance of mobile service providers’ companies could be interesting and the comparison of branchless banking in developed and developing countries can extend understanding for adoption, efficiency, performance and trust.  \\
  
Although most of the studies (82\%) support the argument that DFS improve the performance and profitability of the firm, however, few studies (18\%) doesn't support this argument. Also few studies e.g \citep{Siddik:2016},\citep{Van:2015},\citep{Ceylan:2008},and \citep{Hernando:2007} indicated the time lag of more than two years to get the benefit from such IT investments. This leads to an important aspect missing in the literature that how much should be digitized or what should be the optimal digitization in terms of return (both financial and non-financial) on these investments. Also which factors are affecting this optimal digitization. For example, it could be interesting to investigate whether Technological advancement and customer affinity (awareness to use) are running at the same pace or not. Because if the technology is running at the higher rate than the consumer awareness and utilization, then the adoption by the organizations would likely lead to loss or the lag period to get benefit may increase more.  \\
 
The other important area missing in the literature is the state bonds and lotteries. Previously one could only buy such products at particular bank or store but due to DFS now one can buy these online and even the funds can transfer or withdraw directly from home with internet access. This not only increased sale and customers locally but also globally. It is needed to investigate how DFS are affecting such products and organizations.  
 
 \section{Conclusion}\label{con} 
 This paper systemically reviewed the existing (within last one decade) amount of literature investigating the impact of DFS on firm’s performance, identified the research gaps and outlined seven specific future areas of research. We found only 39 relevant articles out of which only four got citation above 100, furthermore only half of them were published in high ranked journals. We conclude that despite rapid technological advancement in DFS during the last ten years, DFS being the factor affecting firm’s performance didn’t get the attention in academic literature. One of the reason is that almost all the authors focused their research only on banking sector while ignoring others particularly mobile network operators (providing branchless banking) and new non-banking entrants. We also noticed that newer researchers (mainly in developing countries) have often ignored past research and investigated the same issues. This also explains why we have seen such limited progress in last ten years in this area of research.  \\
 
 Overall the research conducted on DFS and bank's performance is promising. This research tends to integrate the other sectors and digital channels that have not been studied so far and need more attention. It is understandable that data availability remained a big challenge due to privacy issues but a collaborative effort from academicians and practitioners could overcome it in future. The constant evolution of technology and complexity of products call for additional research. There are still numerous areas related to DFS and firm's performance that require deeper attention in future. We expect that this literature review and our identified future research areas will provide useful guidance for the research community interested in DFS.

\nocite{*}
\bibliographystyle{unsrtnat}

\end{document}